\documentclass{article} 

\usepackage{amsmath,amssymb,amsthm,mathtools}
\usepackage{booktabs}
\usepackage{graphicx}
\usepackage{xcolor}
\usepackage{hyperref}
\usepackage{url}
\usepackage{algorithm}
\usepackage[noend]{algpseudocode}
\usepackage[capitalise,nameinlink]{cleveref}
\hypersetup{colorlinks=true,linkcolor=blue!60!black,citecolor=blue!60!black,urlcolor=blue!60!black}

\newtheorem{theorem}{Theorem}
\newtheorem{proposition}[theorem]{Proposition}
\newtheorem{lemma}[theorem]{Lemma}
\newtheorem{corollary}[theorem]{Corollary}
\theoremstyle{definition}
\newtheorem{definition}{Definition}
\newtheorem{assumption}{Assumption}
\newtheorem{example}{Example}
\theoremstyle{remark}
\newtheorem{remark}{Remark}
\crefname{assumption}{Assumption}{Assumptions}
\crefname{example}{Example}{Examples}
\crefname{remark}{Remark}{Remarks}
\crefname{definition}{Definition}{Definitions}
\crefname{algorithm}{Algorithm}{Algorithms}

\newcommand{\R}{\mathbb{R}}
\newcommand{\E}{\mathbb{E}}
\renewcommand{\P}{\mathbb{P}}
\newcommand{\Cov}{\operatorname{Cov}}
\newcommand{\Var}{\operatorname{Var}}
\newcommand{\tr}{\operatorname{tr}}
\newcommand{\conv}{\operatorname{conv}}
\newcommand{\supp}{\operatorname{supp}}
\newcommand{\Reg}{\mathrm{Regret}}
\newcommand{\Regn}{\widehat{\mathrm{Regret}}_n}
\newcommand{\chat}{\hat c}
\newcommand{\pis}{\pi^{\star}}
\newcommand{\T}{^{\top}}
\newcommand{\norm}[1]{\lVert #1\rVert}
\newcommand{\opnorm}[1]{\lVert #1\rVert_{\mathrm{op}}}
\usepackage{natbib}
\usepackage[margin=1in]{geometry}

\title{Regret Equals Integrated Covariance:\\Exact Identities for the Value of Information\\in Stochastic Optimization}

\author{Irene Aldridge \\
Cornell Tech, Risk AI Center \\
New York, NY 10044, USA \\
\texttt{irene.aldridge@gmail.com}
}
\date{}
\begin{document}

\maketitle

\begin{abstract}
For a stochastic program $\min_{z\in Z} c\T z + h(z)$ with a random cost vector $c$, regret, or the expected value of perfect information, is usually computed by re-solving the problem scenario by scenario. We show that regret is determined by the covariance between the cost vector and the optimal decision $\pis(c)$. The identity is exact and holds without regularity assumptions once the covariance is integrated along the path from the mean to each scenario. With $c_s=\E[c]+s(c-\E[c])$, $\Reg=-\int_0^1 s^{-1}\Cov(c_s,\pis(c_s))\,ds$. Three consequences follow. First, the plain covariance is always an upper bound: $0\le\Reg\le-\Cov(c,\pis(c))$. Second, when the optimal decision rule is affine, as in unconstrained and equality-constrained (e.g., budget-constrained Markowitz) quadratic programs, regret equals exactly one half of the covariance. For purely linear objectives, an affine rule forces zero regret. Third, the ratio of regret to covariance lies in $[0,1]$ and is governed by the shape of the covariance profile along the path. The noise level at which this profile first becomes positive is a decision-stability margin that tightens the bound. The identity also holds scenario by scenario for any predicted cost, so it applies to linear and combinatorial programs with discontinuous decision rules. We give dimension-free concentration bounds and a central limit theorem for the plug-in estimator. Across 7 problem classes, 8 cost distributions, and the shortest-path benchmark of \citet{elmachtoub2022smart} with $10^6$ draws per configuration, the identities hold to within quadrature error, and the covariance bound holds in every scenario.
\end{abstract}

\section{Introduction}

Decisions are often made from a point estimate of an uncertain cost vector. The loss from doing so, relative to deciding after the costs are revealed, is the expected value of perfect information (EVPI) of stochastic programming \citep{raiffa1961applied,birge2011introduction}, called regret in machine learning \citep{elmachtoub2022smart}. It is usually estimated by sample average approximation (SAA): draw scenarios, solve the problem once per scenario, and average \citep{kleywegt2002sample}. SAA reports how large regret is, but not what generates it.

This paper shows that regret is generated by covariance, the co-movement of the cost vector $c$ and the decision $\pis(c)$ that would be optimal if $c$ were known. The intuition is simple. If decisions did not respond to costs, knowing the costs would be worthless. The more strongly the optimal decision moves against the cost, i.e.\ the more negative $\Cov(c,\pis(c))$, the more information is worth. We make this exact. Writing $\mu=\E[c]$ and $c_s=\mu+s(c-\mu)$ for the cost scaled toward its mean,
\begin{equation}\label{eq:intro}
  \Reg \;=\; -\int_0^1 \frac{\Cov\bigl(c_s,\pis(c_s)\bigr)}{s}\,ds
  \qquad\text{and}\qquad
  0\;\le\;\Reg\;\le\;-\Cov\bigl(c,\pis(c)\bigr).
\end{equation}
Regret is thus the covariance of inputs and outputs, accumulated as uncertainty is switched on from zero ($s=0$) to its actual level ($s=1$). The identity needs only concavity of the optimal value function, so it holds for linear and combinatorial programs, where $\pis$ is piecewise constant and discontinuous.

\paragraph{Contributions.}
\begin{enumerate}
  \item \textbf{Regret equals integrated covariance} (\cref{cor:evpi}). The integral identity \eqref{eq:intro} holds with no assumptions on $\pis$ beyond measurability. It follows from a scenario-wise identity (\cref{thm:pathwise}) that applies to the loss of any predicted cost $\chat$, not only the mean.
  \item \textbf{Covariance is always an upper bound} (\cref{cor:evpi,prop:decomp}). The gap between $-\Cov(c,\pis(c))$ and regret is a residual that we prove is nonnegative.
  \item \textbf{Regret equals exactly half the covariance for affine decision rules} (\cref{thm:affine}). This covers unconstrained and equality-constrained quadratic programs, including budget-constrained Markowitz portfolios. For purely linear objectives, an affine $\pis$ must be constant (\cref{prop:lp-trivial}). The exact covariance formula is thus a quadratic, certainty-equivalence phenomenon \citep{simon1956dynamic,theil1957note}.
  \item \textbf{The regret-to-covariance ratio is a diagnostic} (\cref{prop:shape}). It lies in $[0,1]$, equals $\tfrac12$ for affine rules, lies below $\tfrac12$ when decisions respond to cost perturbations at an accelerating rate, and lies above $\tfrac12$ when they respond early, e.g.\ when the mean sits on a decision boundary. The noise level at which decisions start to respond is a \emph{decision-stability margin} $s^\star$, and it tightens the covariance bound to $\Reg\le(1-s^\star)(-\Cov)$ (\cref{prop:margin}).
  \item \textbf{Estimation} (\cref{thm:concentration,thm:clt}). The plug-in estimator satisfies dimension-free sub-Gaussian bounds for bounded and Gaussian costs, has a nonpositive bias of order $n^{-1/2}$, and is asymptotically normal with variance equal to the variance of the per-scenario loss.
\end{enumerate}

\paragraph{Scope.}
Estimating $\Cov(c,\pis(c))$ requires $\pis(c_i)$ for every scenario, so in general it costs as many solver calls as SAA (\cref{tab:cost}). Savings arise when $\pis$ is affine, where regret is $-\frac12\tr(M\Sigma)$ with no solves, or when an upper bound suffices. The identities are mainly exact and explanatory: they say which co-movements of costs and decisions generate regret.

\section{Related work}

\paragraph{Value of information.} EVPI is classical \citep{raiffa1961applied,howard1966information}. Bounds on EVPI and on the gap between stochastic and mean-value solutions go back to \citet{madansky1960inequalities}, \citet{avriel1970value} and \citet{huang1977sharp}; see \citet{birge2011introduction}. These works bound EVPI via Jensen-type inequalities; our results are equalities in the joint moments of costs and decisions. For SAA see \citet{kleywegt2002sample,shapiro2009lectures}.

\paragraph{Certainty equivalence.} For quadratic criteria with linear dynamics, the optimal decision under uncertainty coincides with the decision under mean values \citep{simon1956dynamic,theil1957note}. \Cref{thm:affine} is a static counterpart for regret: whenever the decision rule is affine, regret is exactly half of the input--output covariance, for every cost distribution with finite second moments.

\paragraph{Predict-then-optimize.} When decisions are made from predicted costs $\chat=\chat(x)$, the realized excess cost is the loss studied by \citet{elmachtoub2022smart}. Its statistical properties are analyzed in \citet{liu2021risk,honguyen2022risk,elbalghiti2023generalization}; see \citet{mandi2024decision,sadana2025survey} for surveys. \Cref{thm:pathwise} applies to this loss for any predictor, and we use the benchmark of \citet{elmachtoub2022smart} only as a testbed for the identities.

\paragraph{Sensitivity analysis.} The supergradient property $\pis(c)\in\partial v(c)$ is the envelope theorem \citep{milgrom2002envelope}. For linear and quadratic programs, $\pis$ is piecewise affine over a polyhedral partition \citep{fiacco1983introduction,bonnans2000perturbation}. We use only concavity of $v$, which holds for every feasible set.

\section{Setup}\label{sec:setup}

Let $Z\subseteq\R^d$ be nonempty and closed, let $h:Z\to\R$ be lower semicontinuous, and write $f(z;c)=c\T z+h(z)$. The case $h\equiv0$ is a linear (or combinatorial) program. $Z$ need not be convex, so integer feasible sets are allowed.

\begin{assumption}[Problem]\label{ass:problem}
For every $c\in\R^d$ the minimum $v(c)=\min_{z\in Z}f(z;c)$ is attained, and $\pis:\R^d\to Z$ is a Borel-measurable selection with $\pis(c)\in\arg\min_{z\in Z}f(z;c)$. Such a selection exists under the stated semicontinuity; see \citet[Thm.~14.37]{rockafellar1998variational}.
\end{assumption}

\begin{assumption}[Growth and moments]\label{ass:growth}
Either (a) $Z$ is bounded, $D:=\sup_{z\in Z}\norm{z}<\infty$, and $\E\norm{c}<\infty$; or (b) $\norm{\pis(c)}\le a+b\norm{c}$ for all $c$ and $\E\norm{c}^2<\infty$.
\end{assumption}

Case (a) covers linear, combinatorial and constrained quadratic programs on bounded sets. Case (b) covers unconstrained and equality-constrained quadratic programs. Let $\mu=\E[c]$ and $\Sigma=\Cov(c)$. For a map $\pi$, we write $\Cov(c,\pi(c)):=\E[(c-\mu)\T\pi(c)]$ for the scalar covariance, i.e.\ the trace of the cross-covariance matrix of $c$ and $\pi(c)$.

\begin{definition}[Regret]\label{def:regret}
The loss of acting on a cost estimate $\chat$ when the realized cost is $c$ is $\ell(\chat,c):=f(\pis(\chat);c)-v(c)\ge0$. Regret, i.e.\ the EVPI of the single-stage program, is
\[
  \Reg := v(\mu)-\E[v(c)] = \E\bigl[\ell(\mu,c)\bigr].
\]
\end{definition}

The second equality holds because $\E[f(\pis(\mu);c)]=\mu\T\pis(\mu)+h(\pis(\mu))=v(\mu)$, so regret is the expected loss of acting on the mean. Throughout, regret is evaluated with the full objective $f$, including $h$. Evaluating only the linear part $c\T z$ for a quadratic program doubles it (\cref{rem:linear-part}).

\begin{lemma}[Value function]\label{lem:value}
Under \cref{ass:problem}, $v$ is concave and finite on $\R^d$, and $\pis(c)$ is a supergradient of $v$ at $c$:
$v(c')\le v(c)+(c'-c)\T\pis(c)$ for all $c,c'$. Under \cref{ass:growth}(a), $v$ is $D$-Lipschitz.
\end{lemma}

\begin{proof}
$v$ is a pointwise minimum of the affine functions $c\mapsto c\T z+h(z)$, $z\in Z$, hence concave, and it is finite by attainment. For any $c'$, $v(c')\le f(\pis(c);c')=v(c)+(c'-c)\T\pis(c)$. If $\norm{z}\le D$ on $Z$, the supergradient inequality applied in both directions gives $|v(c)-v(c')|\le D\norm{c-c'}$.
\end{proof}

\section{Regret equals integrated covariance}\label{sec:identities}

The engine of the paper is an exact statement about concave functions along a segment.

\begin{lemma}[Segment identity]\label{lem:segment}
Fix $a,\Delta\in\R^d$, let $\varphi(s)=v(a+s\Delta)$ and $g(s)=\Delta\T\pis(a+s\Delta)$ for $s\in[0,1]$, and suppose $\sup_{s\in[0,1]}\norm{\pis(a+s\Delta)}<\infty$. Then $g$ is nonincreasing and
\[
  \varphi(1)-\varphi(0)=\int_0^1 g(s)\,ds .
\]
\end{lemma}

\begin{proof}
By \cref{lem:value} at the points $a+s\Delta$ and $a+t\Delta$, $\varphi(t)\le\varphi(s)+(t-s)g(s)$ and $\varphi(s)\le\varphi(t)+(s-t)g(t)$. Adding the two gives $(t-s)(g(s)-g(t))\ge0$, so $g$ is nonincreasing. For $s<t$ the same inequalities give $(t-s)g(t)\le\varphi(t)-\varphi(s)\le(t-s)g(s)$. Summing over a partition $0=s_0<\dots<s_N=1$ sandwiches $\varphi(1)-\varphi(0)$ between a lower and an upper Riemann sum of $g$. Because $g$ is monotone and bounded, it is Riemann integrable, and both sums converge to $\int_0^1g$.
\end{proof}

No differentiability of $v$ and no continuity of $\pis$ is used. For a linear program, $g$ is a step function that jumps whenever the segment crosses a boundary of the normal fan of $\conv Z$.

\begin{theorem}[Scenario-wise identity]\label{thm:pathwise}
Let \cref{ass:problem} hold and let $\chat,c\in\R^d$, $\Delta=c-\chat$, with $\pis$ bounded on the segment $[\chat,c]$, which is automatic under \cref{ass:growth}. Then
\begin{equation}\label{eq:pathwise}
  \ell(\chat,c)=-\int_0^1\Delta\T\bigl(\pis(\chat+s\Delta)-\pis(\chat)\bigr)ds,
  \qquad
  0\le\ell(\chat,c)\le\kappa(\chat,c):=-\Delta\T\bigl(\pis(c)-\pis(\chat)\bigr).
\end{equation}
\end{theorem}

\begin{proof}
First, $f(\pis(\chat);c)=\chat\T\pis(\chat)+h(\pis(\chat))+\Delta\T\pis(\chat)=v(\chat)+g(0)$ with $g$ as in \cref{lem:segment} for $a=\chat$. Second, \cref{lem:segment} gives $v(c)=v(\chat)+\int_0^1g$. Hence $\ell=g(0)-\int_0^1g=-\int_0^1(g(s)-g(0))ds$. Since $g$ is nonincreasing, $g(1)\le g(s)\le g(0)$, which yields $0\le\ell\le g(0)-g(1)=\kappa$.
\end{proof}

We call $\kappa$ the \emph{covariance term}: it is the product of the input error $c-\chat$ with the output change $\pis(c)-\pis(\chat)$. Taking expectations with $\chat=\mu$ turns it into a covariance.

\begin{corollary}[Regret equals integrated covariance]\label{cor:evpi}
Under \cref{ass:problem,ass:growth}, let $c_s:=\mu+s(c-\mu)$. Then
\begin{equation}\label{eq:evpi}
  \Reg=-\int_0^1\E\bigl[(c-\mu)\T\pis(c_s)\bigr]ds=-\int_0^1\frac{\Cov(c_s,\pis(c_s))}{s}\,ds,
\end{equation}
and $0\le\Reg\le-\Cov\bigl(c,\pis(c)\bigr)$. Moreover, the covariance profile $\gamma(s):=-\E[(c-\mu)\T\pis(c_s)]=-\Cov(c_s,\pis(c_s))/s$ is nonnegative and nondecreasing on $(0,1]$, with $\gamma(1)=-\Cov(c,\pis(c))$ and $\gamma(s)\to0$ as $s\downarrow0$ whenever $\pis$ is continuous at $\mu$.
\end{corollary}

\begin{proof}
Apply \cref{thm:pathwise} with $\chat=\mu$ and take expectations; $\E[\ell(\mu,c)]=\Reg$ by \cref{def:regret}. The integrand is bounded by $\norm{c-\mu}\sup_s\norm{\pis(c_s)}$, which is integrable under either branch of \cref{ass:growth}, so Fubini applies. The term $\E[(c-\mu)\T\pis(\mu)]$ is zero. Since $\E[c_s]=\mu$, $\Cov(c_s,\pis(c_s))=\E[(c_s-\mu)\T\pis(c_s)]=s\,\E[(c-\mu)\T\pis(c_s)]$. Monotonicity of $\gamma$ follows from that of $g$ in \cref{lem:segment}, scenario by scenario. Nonnegativity follows from $g(s)\le g(0)$ and $\E[(c-\mu)\T\pis(\mu)]=0$, since $\gamma(s)=-\E[g(s)-g(0)]$ with $a=\mu$. The upper bound is $\E[\kappa(\mu,c)]=\gamma(1)$. The limit at $s\downarrow0$ follows by dominated convergence and $\E[c-\mu]=0$.
\end{proof}

For a contextual prediction $\chat=\chat(x)$, the same argument applies conditionally on $x$. For the Bayes predictor $\chat(x)=\E[c\mid x]$, the expected loss equals $-\int_0^1s^{-1}\E[\Cov(c_s,\pis(c_s)\mid x)]ds$ with $c_s=\chat(x)+s(c-\chat(x))$, and it is bounded above by the expected conditional covariance $-\E[\Cov(c,\pis(c)\mid x)]$.

\paragraph{The residual.}
Adding and subtracting $\mu\T\E[\pis(c)]$ in \cref{def:regret} relates regret to the plain covariance directly.

\begin{proposition}[Covariance decomposition]\label{prop:decomp}
$\Reg=-\Cov(c,\pis(c))-R-H$, where $R:=\mu\T(\E[\pis(c)]-\pis(\mu))$ and $H:=\E[h(\pis(c))]-h(\pis(\mu))$. By \cref{cor:evpi}, $R+H\ge0$. If $h\equiv0$ and $Z$ is bounded, then $R\ge0$ on its own, since $\E[\pis(c)]\in\overline{\conv}\,Z$ and $\pis(\mu)$ minimizes $\mu\T z$ over $\conv Z$.
\end{proposition}

The residual is never negative, so covariance always overstates regret. \Cref{cor:evpi} replaces the residual with an exact integral.

\begin{example}[A one-dimensional LP]\label{ex:lp}
Let $Z=[0,1]$, $h\equiv0$, $c\sim\mathrm{Unif}[-2,1]$, so $\mu=-\tfrac12$ and $\pis(c)=\mathbf 1\{c<0\}$. Directly, $\Reg=v(\mu)-\E\min(c,0)=-\tfrac12+\tfrac23=\tfrac16$, $-\Cov(c,\pis(c))=\tfrac13$, and $R=\tfrac16$. With $\Delta=c-\mu\sim\mathrm{Unif}[-\tfrac32,\tfrac32]$, the profile is $\gamma(s)=-\E[\Delta\,\mathbf 1\{\Delta<1/(2s)\}]$. It equals $0$ for $s\le\tfrac13$ and $\tfrac38-\tfrac1{24s^2}$ for $s>\tfrac13$, and $\int_0^1\gamma=\tfrac16$ exactly. Decisions do not respond at all until the scaled perturbation can cross the boundary at $c=0$.
\end{example}

\paragraph{Affine decision rules.} When $\pis$ is affine the integral is evaluated in closed form, and covariance determines regret exactly.

\begin{theorem}[Regret is exactly half the covariance]\label{thm:affine}
Let $K$ be the closed convex hull of $\supp(P_c)$ and suppose $\pis(c)=Ac+b$ for all $c\in K$. Under \cref{ass:problem,ass:growth},
\[
  \Reg=-\tfrac12\tr(A\Sigma)=-\tfrac12\Cov\bigl(c,\pis(c)\bigr).
\]
\end{theorem}

\begin{proof}
$\mu\in K$ and $K$ is convex, so $c_s\in K$ and $\pis(c_s)=A\mu+b+sA(c-\mu)$. In \eqref{eq:evpi}, $\E[(c-\mu)\T(A\mu+b)]=0$, so $\Reg=-\int_0^1s\,\E[(c-\mu)\T A(c-\mu)]ds=-\tfrac12\tr(A\Sigma)$. Finally, $\Cov(c,Ac+b)=\tr(A\Sigma)$.
\end{proof}

The result requires no distributional assumption beyond finite second moments: heavy-tailed, discrete, skewed and multimodal costs all give a ratio of exactly $\tfrac12$ (\cref{tab:synthetic}).

\begin{corollary}[Quadratic programs]\label{cor:qp}
Let $h(z)=\tfrac\lambda2z\T Qz$ with $Q\succ0$ and $\lambda>0$.
(i) If $Z=\R^d$, then $\pis(c)=-\lambda^{-1}Q^{-1}c$ and $\Reg=\tr(Q^{-1}\Sigma)/(2\lambda)$.
(ii) If $Z=\{z:Az=b\}$ with $A$ of full row rank (e.g.\ the Markowitz budget $\mathbf1\T z=1$), then $\pis(c)=Mc+m_0$ with $M=-\lambda^{-1}\bigl(Q^{-1}-Q^{-1}A\T(AQ^{-1}A\T)^{-1}AQ^{-1}\bigr)$, and $\Reg=-\tfrac12\tr(M\Sigma)$.
(iii) With inequality constraints (e.g.\ no short sales), $\pis$ is a $Q$-weighted projection. It is Lipschitz continuous and piecewise affine but in general not affine, so only \cref{cor:evpi} applies.
\end{corollary}

Derivations are in \cref{app:qp}. In case (ii), regret is available in closed form from $\Sigma$ alone, with no optimization solves.

\begin{remark}[Linear part only]\label{rem:linear-part}
For a quadratic program, the quantity $\mu\T\pis(\mu)-\E[c\T\pis(c)]$, which omits $h$, equals $-\Cov(c,\pis(c))-R$, i.e.\ $\tr(Q^{-1}\Sigma)/\lambda$ in case (i). That is twice the EVPI. Measuring regret with the full objective $f$ is what makes it the expected loss of acting on the mean.
\end{remark}

\begin{proposition}[Affine rules are trivial for linear objectives]\label{prop:lp-trivial}
Suppose $h\equiv0$, $K$ has nonempty interior, and $\pis(c)=Ac+b$ on $K$. Then $A=0$, so $\pis$ is constant on $K$ and $\Reg=0$.
\end{proposition}

\begin{proof}
A concave function is differentiable almost everywhere on the interior of its domain \citep[Thm.~25.5]{rockafellar1970convex}, and where it is differentiable its gradient is its unique supergradient, so $\nabla v(c)=Ac+b$ for almost every $c\in\operatorname{int}K$. With $h\equiv0$, $v(c)=c\T\pis(c)=c\T Ac+b\T c$ on $K$, whose gradient is $(A+A\T)c+b$. Equating the two gives $A\T c=0$ on a set of positive measure, so $A=0$.
\end{proof}

The exact half-covariance formula is therefore a feature of curved objectives. For linear and combinatorial programs, the exact object is the integral \eqref{eq:evpi}, and the endpoint covariance is a strict upper bound whenever regret is positive.

\section{What the regret-to-covariance ratio reveals}\label{sec:shape}

By \cref{cor:evpi}, $\Reg=\int_0^1\gamma(s)\,ds$ and $-\Cov(c,\pis(c))=\gamma(1)$, where $\gamma$ is the nondecreasing covariance profile. Their ratio
\[
  \rho:=\frac{\Reg}{-\Cov(c,\pis(c))}=\frac{\int_0^1\gamma(s)\,ds}{\gamma(1)}\in[0,1]
\]
is the average height of the profile relative to its endpoint. It summarizes \emph{when}, along the path from certainty to full uncertainty, decisions start responding to costs.

\begin{proposition}[Shape of the covariance profile]\label{prop:shape}
Suppose $\gamma(1)>0$.
(i) $\rho\in[0,1]$.
(ii) If $\pis$ is affine on $K$, then $\gamma(s)=s\,\gamma(1)$ and $\rho=\tfrac12$.
(iii) If $\gamma$ is convex on $[0,1]$ with $\gamma(0^+)=0$, then $\rho\le\tfrac12$; if $\gamma$ is concave, then $\rho\ge\tfrac12$.
(iv) $\rho=1$ if and only if $\gamma$ is constant on $(0,1]$, i.e.\ the covariance per unit of noise is fully attained at infinitesimal noise.
\end{proposition}

\begin{proof}
(i) $0\le\gamma\le\gamma(1)$ by monotonicity. (ii) Follows from the proof of \cref{thm:affine}. (iii) A convex function with $\gamma(0^+)=0$ lies below its chord, $\gamma(s)\le s\gamma(1)$, so $\int_0^1\gamma\le\frac12\gamma(1)$. The concave case reverses the inequality. (iv) Since $\gamma\le\gamma(1)$ and $\gamma$ is monotone, $\int_0^1\gamma=\gamma(1)$ forces $\gamma=\gamma(1)$ on $(0,1]$.
\end{proof}

A convex profile, $\rho<\frac12$, means decisions are insensitive to small cost perturbations and switch only when perturbations are large. This is typical when $\mu$ lies deep inside one region of the decision partition, as in \cref{ex:lp} up to $s=\tfrac13$. A profile that rises early, $\rho>\frac12$, means decisions switch already under small perturbations. The extreme case $\rho\to1$ occurs when $\mu$ lies on a decision boundary with probability mass on both sides. A single number from $n$ solves, the covariance, therefore bounds regret within a factor that the shape of $\gamma$ explains, and $\gamma$ can be estimated from the same samples at $K$ additional solves per scenario.

\paragraph{Decision stability.}
Since $\gamma\ge0$ is nondecreasing, the set where it vanishes is an interval starting at $0$. Its right endpoint
$s^\star:=\sup\{s\in[0,1]:\gamma(s)=0\}$
is the largest fraction of the actual uncertainty under which the certainty-equivalent decision $\pis(\mu)$ is, on average, unaffected. We call it the \emph{decision-stability margin}.

\begin{proposition}[Stability margin]\label{prop:margin}
(i) $\Reg=\int_{s^\star}^1\gamma(s)\,ds\le(1-s^\star)\bigl(-\Cov(c,\pis(c))\bigr)$, so $\rho\le1-s^\star$.
(ii) If $\norm{c-\mu}\le r$ almost surely and $\pis(c')=\pis(\mu)$ whenever $\norm{c'-\mu}<r'$, then $s^\star\ge\min\{r'/r,1\}$.
\end{proposition}

\begin{proof}
(i) $\gamma=0$ on $[0,s^\star)$ and $\gamma\le\gamma(1)$. (ii) For $s<r'/r$, $\norm{c_s-\mu}=s\norm{c-\mu}<r'$, so $\pis(c_s)=\pis(\mu)$ and $\gamma(s)=-\E[(c-\mu)\T\pis(\mu)]=0$.
\end{proof}

Thus $r'/r$, the distance from the mean to the nearest decision boundary measured in units of the noise, certifies stability and shrinks the covariance bound. In \cref{ex:lp}, $s^\star=\tfrac13$ and $\rho=\tfrac12\le\tfrac23$. In the box LP of \cref{sec:exp-synthetic}, the mean lies on a boundary, so $s^\star=0$ and $\rho\approx1$. For costs with unbounded support, $s^\star$ is typically $0$, and the small-$s$ growth of $\gamma$ plays the same role. This is stability of the \emph{decision} map $c\mapsto\pis(c)$ around $\mu$. It is distinct from algorithmic stability of a learned predictor with respect to its training data \citep{bousquet2002stability}, about which integrated covariance says nothing.

\section{Bounds and estimation}\label{sec:estimation}

\begin{theorem}[Distribution-free bound]\label{thm:bound}
Under \cref{ass:growth}(a), $0\le\Reg\le\min\bigl\{-\Cov(c,\pis(c)),\;D\,\E\norm{c-\mu}\bigr\}\le D\sqrt{\tr\Sigma}$.
\end{theorem}

\begin{proof}
The first term is from \cref{cor:evpi}. For the second, $\Reg=\E[v(\mu)-v(c)]\le D\,\E\norm{c-\mu}$ by \cref{lem:value}, and $\E\norm{c-\mu}\le\sqrt{\tr\Sigma}$ by Jensen's inequality.
\end{proof}

Neither bound dominates the other. The covariance bound reflects the decision rule but needs $\pis(c_i)$ for every sample. The radius bound needs no solves; \cref{app:examples} lists $D$ for common combinatorial feasible sets.

Given i.i.d.\ samples $c_1,\dots,c_n$ with mean $\bar c_n$, the plug-in estimator is $\Regn=v(\bar c_n)-\frac1n\sum_iv(c_i)$.

\begin{theorem}[Concentration and bias]\label{thm:concentration}
Under \cref{ass:growth}(a):
(i) if $\supp(P_c)$ has diameter $\delta$, then $\P(|\Regn-\E\Regn|\ge u)\le2\exp(-nu^2/(2D^2\delta^2))$;
(ii) if $c\sim N(\mu,\Sigma)$, then $\P(|\Regn-\E\Regn|\ge u)\le2\exp(-nu^2/(8D^2\opnorm{\Sigma}))$;
(iii) in either case, $-D\sqrt{\tr\Sigma/n}\le\E\Regn-\Reg\le0$.
\end{theorem}

Part (ii) does not require bounded support and does not depend on the dimension except through $\opnorm\Sigma$. It follows from Gaussian concentration of Lipschitz functions \citep{tsirelson1976norms}, part (i) from bounded differences \citep{mcdiarmid1989method}, and part (iii) from concavity; proofs are in \cref{app:proofs}.

\begin{theorem}[Asymptotic normality]\label{thm:clt}
Under \cref{ass:growth}(a) with $\E\norm{c}^2<\infty$, suppose $v$ is differentiable at $\mu$. For compact $Z$ and continuous $h$, this is equivalent to uniqueness of the minimizer at $\mu$. Then
$\sqrt n(\Regn-\Reg)\Rightarrow N\bigl(0,\Var(\ell(\mu,c))\bigr)$.
\end{theorem}

The asymptotic variance is the variance of the per-scenario loss of acting on the mean, so a confidence interval needs only the sample variance of $c_i\T\pis(\bar c_n)+h(\pis(\bar c_n))-v(c_i)$. If $v$ has a kink at $\mu$, the limit of $\sqrt n(v(\bar c_n)-v(\mu))$ is $\min_{w\in\partial v(\mu)}w\T G$ with $G\sim N(0,\Sigma)$, which is not Gaussian.

\paragraph{Computing integrated covariance.}
\Cref{alg:intcov} turns \cref{cor:evpi,thm:pathwise} into a procedure. It accepts pairs $(\chat_i,c_i)$ of cost estimates and realizations. Setting $\chat_i=\bar c_n$ for all $i$ gives regret (EVPI); setting $\chat_i=\chat(x_i)$ gives the expected loss of a predictor. For each quadrature node $s_k$, it evaluates the decision at the scaled cost $\chat_i+s_k\Delta_i$ and records the covariance profile $\hat\gamma_k$. Regret is the weighted sum of the profile, the covariance bound is its value at $s=1$, $\hat\rho$ is their ratio, and $\hat s^\star$ is the last node at which the profile is still zero.

\begin{algorithm}[t]
\caption{Integrated covariance: regret, covariance bound and profile from samples}
\label{alg:intcov}
\begin{algorithmic}[1]
\Require Realized costs $c_1,\dots,c_n$; cost estimates $\chat_1,\dots,\chat_n$ (default $\chat_i=\bar c_n:=\frac1n\sum_jc_j$); an oracle returning $\pis(c)\in\arg\min_{z\in Z}c\T z+h(z)$; quadrature nodes $0<s_1<\dots<s_K\le1$ with weights $w_1,\dots,w_K$ on $[0,1]$
\Ensure Regret estimate $\widehat{\Reg}$, covariance bound $\hat\gamma(1)$, profile $(\hat\gamma_k)_{k=1}^K$, ratio $\hat\rho$, stability margin $\hat s^\star$
\For{$i=1,\dots,n$}
  \State $\Delta_i\gets c_i-\chat_i$ \Comment{input error}
  \State $\hat z_i\gets\pis(\chat_i)$ \Comment{decision at the estimate; one solve if $\chat_i\equiv\bar c_n$}
\EndFor
\For{$k=1,\dots,K$} \Comment{switch uncertainty on gradually}
  \For{$i=1,\dots,n$}
    \State $z_{ik}\gets\pis(\chat_i+s_k\Delta_i)$ \Comment{decision at the scaled cost}
  \EndFor
  \State $\hat\gamma_k\gets-\frac1n\sum_{i=1}^n\Delta_i\T\bigl(z_{ik}-\hat z_i\bigr)$ \Comment{covariance profile at $s_k$}
\EndFor
\State $\hat\gamma(1)\gets-\frac1n\sum_{i=1}^n\Delta_i\T\bigl(\pis(c_i)-\hat z_i\bigr)$ \Comment{covariance bound; reuse $z_{iK}$ if $s_K=1$}
\State $\widehat{\Reg}\gets\sum_{k=1}^Kw_k\hat\gamma_k$ \Comment{integrated covariance}
\State $\hat\rho\gets\widehat{\Reg}/\hat\gamma(1)$
\State $\hat s^\star\gets\max\bigl(\{0\}\cup\{s_k:\hat\gamma_k=0\}\bigr)$ \Comment{decision-stability margin (\cref{prop:margin})}
\State \textbf{check:} $0\le\hat\gamma_1\le\dots\le\hat\gamma_K\le\hat\gamma(1)$ \Comment{guaranteed by \cref{lem:segment}; a violation flags a solver error}
\State \Return $\widehat{\Reg},\ \hat\gamma(1),\ (\hat\gamma_k)_k,\ \hat\rho,\ \hat s^\star$
\end{algorithmic}
\end{algorithm}

\Cref{alg:intcov} uses $nK+n+1$ solver calls; when the costs are scaled toward their mean, the decisions $\hat z_i$ require a single solve. Under the empirical distribution of the sample, the output equals the SAA estimate $\Regn$ up to quadrature error, and the monotonicity check holds exactly, scenario by scenario. When $\pis$ is piecewise constant (linear and combinatorial programs), the profile is a step function, so we recommend the midpoint rule ($s_k=(k-\frac12)/K$, $w_k=1/K$), whose error is $O(1/K)$. When $\pis$ is continuous (quadratic programs), Gauss--Legendre nodes converge much faster. If $\pis$ is affine, a single node suffices, because the profile is linear. The $nK$ solves are independent and can be batched; for problems with a modest number of extreme points, such as the $5\times5$ grid with its 70 paths, all of them reduce to one matrix product and an argmin.

\begin{table}[t]
\caption{Solver calls and exactness of regret estimators from $n$ samples. $K$ is the number of quadrature nodes in \eqref{eq:evpi}.}
\label{tab:cost}
\centering
\small
\begin{tabular}{lll}
\toprule
Estimator & Solver calls & Target \\
\midrule
SAA: $v(\bar c_n)-\frac1n\sum_iv(c_i)$ & $n+1$ & Regret (consistent) \\
Covariance: $-\widehat\Cov(c,\pis(c))$ & $n$ & Upper bound; exactly $2\times$ regret if $\pis$ affine \\
Integrated covariance (\cref{alg:intcov}) & $nK+n+1$ & Regret, bound, profile (to quadrature) \\
Closed form $-\frac12\tr(M\widehat\Sigma)$, affine $\pis$ & $0$ (one factorization) & Regret (exact) \\
Radius bound $D\sqrt{\tr\widehat\Sigma}$ & $0$ & Upper bound \\
\bottomrule
\end{tabular}
\end{table}

\section{Experiments}\label{sec:experiments}

All code is in the supplementary material. Optimal decisions are computed exactly: by enumerating the 70 paths of the $5\times5$ grid, by dynamic programming, by closed forms, or by bisection for the simplex-constrained QP. Details are in \cref{app:details}.

\subsection{Identities across problems and cost distributions}\label{sec:exp-synthetic}

We test 7 problem classes in $d=10$ (40 for the shortest path) against 8 cost distributions: correlated Gaussian, uniform, Student-$t_5$, log-normal, gamma, Poisson, a bimodal mixture and a two-point distribution, with $n=20{,}000$ samples each. Because the identities hold for every distribution, we evaluate them under the empirical distribution of the sample with \cref{alg:intcov}, so any error reflects quadrature or floating point only.

\begin{table}[t]
\caption{Identity checks over 8 cost distributions per problem. The $\rho$ column gives the range of $\Reg/(-\Cov)$ over the distributions. The error columns give the maximum relative error of \cref{prop:decomp} and of the integral \eqref{eq:evpi} (48 Gauss--Legendre nodes). The last column counts the cases in which $0\le\Reg\le-\Cov$ holds. In the box LP the threshold equals the mean, which puts $\mu$ on a decision boundary.}
\label{tab:synthetic}
\centering
\small
\begin{tabular}{lcccc}
\toprule
Problem & $\rho=\Reg/(-\Cov)$ & Decomposition err. & Integral err. & Bound holds \\
\midrule
QP, unconstrained & 0.500--0.500 & $1{\times}10^{-15}$ & $9{\times}10^{-16}$ & 8/8 \\
QP, budget only & 0.500--0.500 & $3{\times}10^{-16}$ & $4{\times}10^{-16}$ & 8/8 \\
QP, no short sales & 0.550--0.705 & $5{\times}10^{-16}$ & $9{\times}10^{-6}$ & 8/8 \\
LP, simplex & 0.410--0.791 & $3{\times}10^{-16}$ & $1{\times}10^{-2}$ & 8/8 \\
LP, box & 0.989--0.995 & $4{\times}10^{-16}$ & $5{\times}10^{-4}$ & 8/8 \\
LP, polytope ($K{=}40$) & 0.384--0.719 & $3{\times}10^{-16}$ & $8{\times}10^{-5}$ & 8/8 \\
Shortest path $5{\times}5$ & 0.383--0.760 & $6{\times}10^{-16}$ & $2{\times}10^{-4}$ & 8/8 \\
\bottomrule
\end{tabular}
\end{table}

\begin{figure}[t]
\centering
\includegraphics[width=\linewidth]{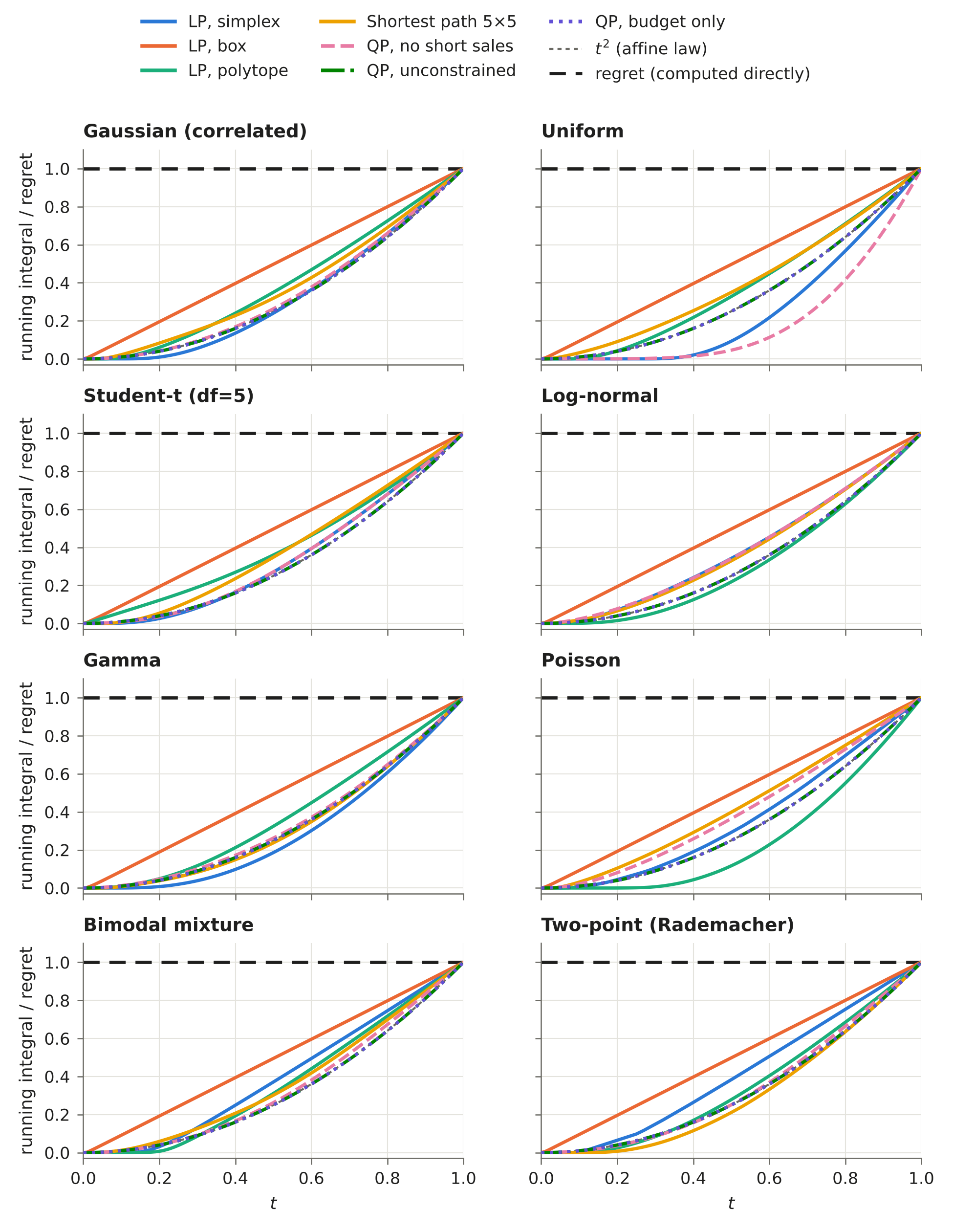}
\caption{Running integral of covariance, $\int_0^t\gamma(s)ds$, divided by regret, for 7 problems (lines) and 8 cost distributions (panels). Every curve reaches the directly computed regret (dashed line at 1) at $t=1$. The dotted $t^2$ curve is the law for affine decision rules.}
\label{fig:curves}
\end{figure}

\Cref{tab:synthetic} confirms each prediction. For both affine QPs, $\rho=0.500$ to three decimals for every distribution, including heavy-tailed, discrete and bimodal costs (\cref{thm:affine}). For every other problem $\rho\in(0,1)$, and the box LP, whose mean sits on a decision boundary ($s^\star=0$), has $\rho\approx0.99$ (\cref{prop:shape,prop:margin}). The integral identity holds to floating-point precision for the affine QPs and to quadrature accuracy elsewhere. The largest error ($10^{-2}$, a two-point distribution with a coarse step-shaped profile) falls to $7\times10^{-4}$ with 400 midpoint nodes.
\Cref{fig:curves} plots the running integral $\int_0^t\gamma(s)\,ds$ normalized by regret. Affine rules trace $t^2$ exactly. Curves below $t^2$ correspond to convex profiles ($\rho<\tfrac12$), and curves above $t^2$ to profiles that rise early ($\rho>\tfrac12$).

\subsection{Contextual costs: the shortest-path benchmark}\label{sec:exp-eg}

We next test the scenario-wise identity (\cref{thm:pathwise}) when decisions are made from feature-based cost predictions. We follow \citet{elmachtoub2022smart}: a $5\times5$ grid with $d=40$ edges, $p=5$ features $x\sim N(0,I_5)$, $B^\star\in\{0,1\}^{40\times5}$ with Bernoulli$(0.5)$ entries, and costs $c_j=[(p^{-1/2}(B^\star x)_j+3)^{\deg}+1]\varepsilon_j$, $\varepsilon_j\sim\mathrm{Unif}[0.5,1.5]$, for $\deg=1,\dots,8$. Larger $\deg$ makes costs more nonlinear in $x$. We consider two predictors: the Bayes predictor $\chat(x)=\E[c\mid x]$, for which $\E[\kappa]$ is the expected conditional covariance $-\E[\Cov(c,\pis(c)\mid x)]$; and least squares on $(1,x)$, trained on 1{,}000 samples, which is misspecified for $\deg>1$. For each configuration we draw $10^6$ scenarios and compute, scenario by scenario, the loss $\ell$, the covariance term $\kappa$, and the integral in \eqref{eq:pathwise}, using \cref{alg:intcov} with $\chat_i=\chat(x_i)$ and 50 midpoint nodes.

\begin{table}[t]
\caption{Shortest path, $10^6$ draws per cell. Regret is $\E\ell$; Cov.\ is $\E\kappa$, the (conditional) covariance term; $\rho=\E\ell/\E\kappa$; Err.\ is $|\E\ell-\text{integral}|/\E\ell$.}
\label{tab:eg}
\centering
\small
\setlength{\tabcolsep}{4.5pt}
\begin{tabular}{c cccc cccc}
\toprule
& \multicolumn{4}{c}{Bayes predictor $\E[c\mid x]$} & \multicolumn{4}{c}{Least-squares predictor} \\
\cmidrule(lr){2-5}\cmidrule(lr){6-9}
deg & Regret & Cov. & $\rho$ & Err. & Regret & Cov. & $\rho$ & Err. \\
\midrule
1 & 3.89 & 5.46 & 0.71 & $6{\times}10^{-5}$ & 3.92 & 5.53 & 0.71 & $5{\times}10^{-5}$ \\
2 & 6.37 & 11.08 & 0.58 & $2{\times}10^{-5}$ & 6.54 & 11.50 & 0.57 & $2{\times}10^{-5}$ \\
3 & 13.96 & 26.21 & 0.53 & $2{\times}10^{-5}$ & 15.51 & 29.56 & 0.52 & $2{\times}10^{-5}$ \\
4 & 37.51 & 72.02 & 0.52 & $4{\times}10^{-5}$ & 49.08 & 99.37 & 0.49 & $5{\times}10^{-6}$ \\
5 & 114 & 219 & 0.52 & $3{\times}10^{-5}$ & 182 & 391 & 0.47 & $2{\times}10^{-5}$ \\
6 & 383 & 723 & 0.53 & $6{\times}10^{-6}$ & 720 & 1{,}572 & 0.46 & $6{\times}10^{-5}$ \\
7 & 1{,}394 & 2{,}565 & 0.54 & $2{\times}10^{-5}$ & 3{,}249 & 7{,}380 & 0.44 & $1{\times}10^{-4}$ \\
8 & 5{,}462 & 9{,}684 & 0.56 & $1{\times}10^{-5}$ & 14{,}849 & 36{,}371 & 0.41 & $2{\times}10^{-4}$ \\
\bottomrule
\end{tabular}
\end{table}

\begin{figure}[t]
\centering
\includegraphics[width=\linewidth]{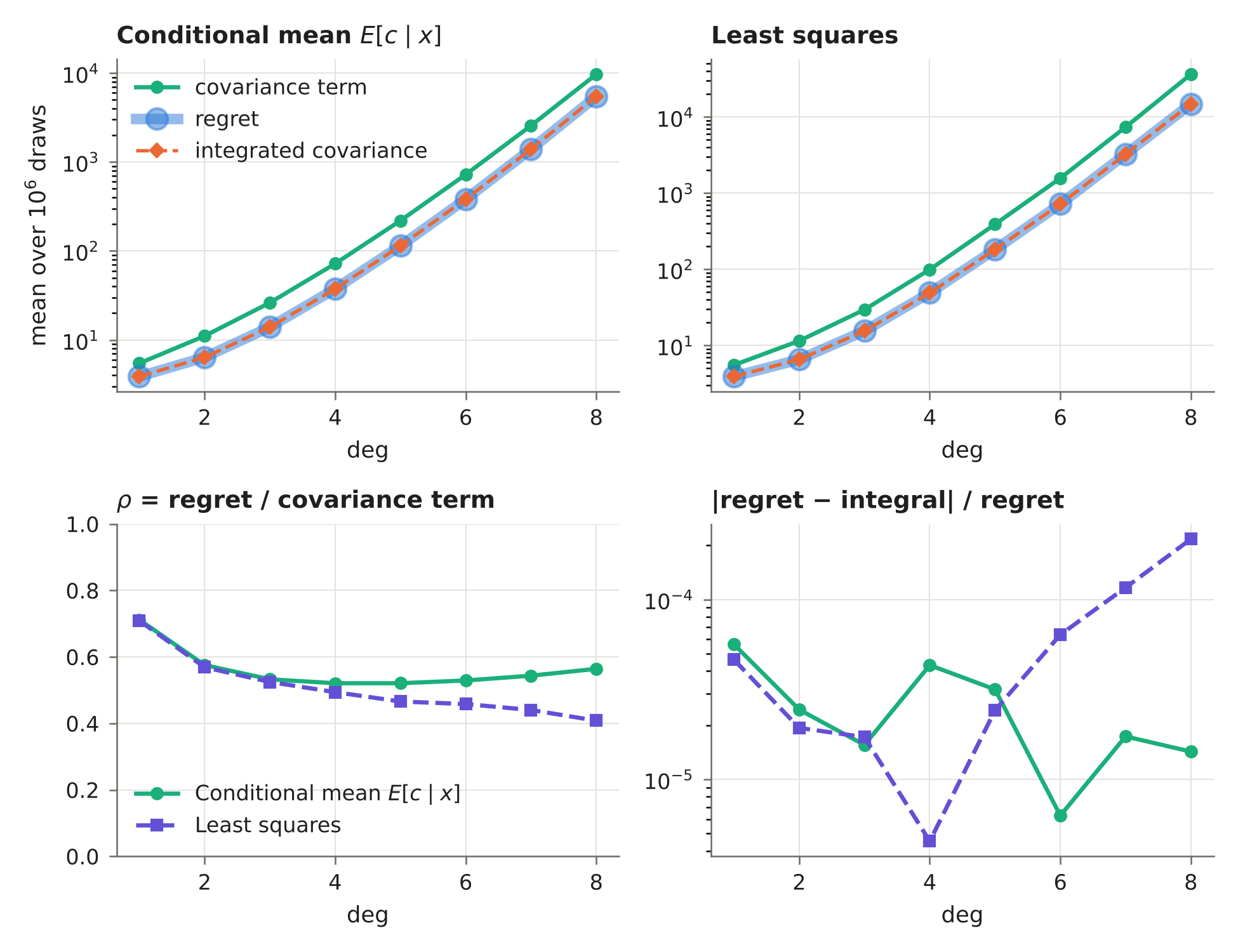}
\caption{Shortest path, $10^6$ draws per point. Left two panels: regret, integrated covariance and covariance term for each predictor (log scale). Third panel: $\rho=$ regret$/$covariance term. Right: relative error of the integral identity.}
\label{fig:sp}
\end{figure}

In \cref{tab:eg,fig:sp}, integrated covariance matches regret to within $2.2\times10^{-4}$ relative error in all 16 configurations, across four orders of magnitude of regret, and $0\le\ell\le\kappa$ held in every one of the $1.6\times10^7$ scenarios (\cref{thm:pathwise}). Because the identity holds scenario by scenario, the running means of regret and of the integral coincide for every number of draws, including a single draw (\cref{fig:convergence}, \cref{app:figures}). For the Bayes predictor, $\rho$ falls from $0.71$ at $\deg=1$ to about $0.52$, near the affine value $\tfrac12$. For the misspecified least-squares predictor it falls steadily to $0.41$, plausibly because its biased errors add a cross-moment to the covariance term that does not translate into loss. The plain covariance thus stays within a factor of about two of regret, and the integrated covariance is exact for both predictors.

\section{Limitations and conclusion}\label{sec:conclusion}

The identities are exact but not, in general, cheaper than SAA. The covariance needs one solve per scenario, and the integrated covariance needs $K$. Savings arise for affine decision rules, where regret is $-\frac12\tr(M\Sigma)$ in closed form, and for the radius bound. The Lipschitz-based estimation results need a bounded feasible set. The ratio diagnostic of \cref{prop:shape} gives a qualitative reading of $\rho$ but not a sharp bound for a given problem class.

Within these limits, the picture is simple. Regret is the covariance between costs and optimal decisions, integrated over the level of uncertainty. The plain covariance is always an upper bound. It equals exactly twice the regret when the decision rule is affine, a case that includes budget-constrained Markowitz portfolios. For affine rules with purely linear objectives, both are zero. The noise level at which the covariance profile first becomes positive measures the stability of the decision and sharpens the bound. Natural next steps are sharp bounds on $\rho$ from the geometry of the normal fan of $Z$, and closed forms for $\gamma$ for Gaussian costs, where Stein's identity links $\Cov(c_s,\pis(c_s))$ to the expected Jacobian of $\pis$.

\bibliography{references}

@book{raiffa1961applied,
  author    = {Raiffa, Howard and Schlaifer, Robert},
  title     = {Applied Statistical Decision Theory},
  publisher = {Division of Research, Graduate School of Business Administration, Harvard University},
  address   = {Boston, MA},
  year      = {1961}
}

@article{howard1966information,
  author  = {Howard, Ronald A.},
  title   = {Information Value Theory},
  journal = {IEEE Transactions on Systems Science and Cybernetics},
  volume  = {2},
  number  = {1},
  pages   = {22--26},
  year    = {1966}
}

@article{madansky1960inequalities,
  author  = {Madansky, Albert},
  title   = {Inequalities for Stochastic Linear Programming Problems},
  journal = {Management Science},
  volume  = {6},
  number  = {2},
  pages   = {197--204},
  year    = {1960}
}

@article{avriel1970value,
  author  = {Avriel, Mordecai and Williams, A. C.},
  title   = {The Value of Information and Stochastic Programming},
  journal = {Operations Research},
  volume  = {18},
  number  = {5},
  pages   = {947--954},
  year    = {1970}
}

@article{huang1977sharp,
  author  = {Huang, C. C. and Vertinsky, Ilan and Ziemba, William T.},
  title   = {Sharp Bounds on the Value of Perfect Information},
  journal = {Operations Research},
  volume  = {25},
  number  = {1},
  pages   = {128--139},
  year    = {1977}
}

@book{birge2011introduction,
  author    = {Birge, John R. and Louveaux, Fran{\c{c}}ois},
  title     = {Introduction to Stochastic Programming},
  edition   = {2nd},
  publisher = {Springer},
  address   = {New York, NY},
  year      = {2011}
}

@book{shapiro2009lectures,
  author    = {Shapiro, Alexander and Dentcheva, Darinka and Ruszczy{\'n}ski, Andrzej},
  title     = {Lectures on Stochastic Programming: Modeling and Theory},
  publisher = {SIAM},
  address   = {Philadelphia, PA},
  year      = {2009}
}

@article{kleywegt2002sample,
  author  = {Kleywegt, Anton J. and Shapiro, Alexander and Homem-de-Mello, Tito},
  title   = {The Sample Average Approximation Method for Stochastic Discrete Optimization},
  journal = {SIAM Journal on Optimization},
  volume  = {12},
  number  = {2},
  pages   = {479--502},
  year    = {2002}
}

@article{simon1956dynamic,
  author  = {Simon, Herbert A.},
  title   = {Dynamic Programming under Uncertainty with a Quadratic Criterion Function},
  journal = {Econometrica},
  volume  = {24},
  number  = {1},
  pages   = {74--81},
  year    = {1956}
}

@article{theil1957note,
  author  = {Theil, Henri},
  title   = {A Note on Certainty Equivalence in Dynamic Planning},
  journal = {Econometrica},
  volume  = {25},
  number  = {2},
  pages   = {346--349},
  year    = {1957}
}

@article{elmachtoub2022smart,
  author  = {Elmachtoub, Adam N. and Grigas, Paul},
  title   = {Smart ``Predict, then Optimize''},
  journal = {Management Science},
  volume  = {68},
  number  = {1},
  pages   = {9--26},
  year    = {2022}
}

@inproceedings{liu2021risk,
  author    = {Liu, Heyuan and Grigas, Paul},
  title     = {Risk Bounds and Calibration for a Smart Predict-then-Optimize Method},
  booktitle = {Advances in Neural Information Processing Systems},
  volume    = {34},
  year      = {2021}
}

@article{elbalghiti2023generalization,
  author  = {El Balghiti, Othman and Elmachtoub, Adam N. and Grigas, Paul and Tewari, Ambuj},
  title   = {Generalization Bounds in the Predict-then-Optimize Framework},
  journal = {Mathematics of Operations Research},
  volume  = {48},
  number  = {4},
  year    = {2023}
}

@article{honguyen2022risk,
  author  = {Ho-Nguyen, Nam and K{\i}l{\i}n{\c{c}}-Karzan, Fatma},
  title   = {Risk Guarantees for End-to-End Prediction and Optimization Processes},
  journal = {Management Science},
  volume  = {68},
  number  = {12},
  year    = {2022}
}

@article{mandi2024decision,
  author  = {Mandi, Jayanta and Kotary, James and Berden, Senne and Mulamba, Maxime and Bucarey, Victor and Guns, Tias and Fioretto, Ferdinando},
  title   = {Decision-Focused Learning: Foundations, State of the Art, Benchmark and Future Opportunities},
  journal = {Journal of Artificial Intelligence Research},
  volume  = {80},
  year    = {2024}
}

@article{sadana2025survey,
  author  = {Sadana, Utsav and Chenreddy, Abhilash and Delage, Erick and Forel, Alexandre and Frejinger, Emma and Vidal, Thibaut},
  title   = {A Survey of Contextual Optimization Methods for Decision-Making under Uncertainty},
  journal = {European Journal of Operational Research},
  volume  = {320},
  number  = {2},
  pages   = {271--289},
  year    = {2025}
}

@book{rockafellar1970convex,
  author    = {Rockafellar, R. Tyrrell},
  title     = {Convex Analysis},
  publisher = {Princeton University Press},
  address   = {Princeton, NJ},
  year      = {1970}
}

@book{rockafellar1998variational,
  author    = {Rockafellar, R. Tyrrell and Wets, Roger J.-B.},
  title     = {Variational Analysis},
  publisher = {Springer},
  address   = {Berlin},
  year      = {1998}
}

@article{milgrom2002envelope,
  author  = {Milgrom, Paul and Segal, Ilya},
  title   = {Envelope Theorems for Arbitrary Choice Sets},
  journal = {Econometrica},
  volume  = {70},
  number  = {2},
  pages   = {583--601},
  year    = {2002}
}

@book{fiacco1983introduction,
  author    = {Fiacco, Anthony V.},
  title     = {Introduction to Sensitivity and Stability Analysis in Nonlinear Programming},
  publisher = {Academic Press},
  address   = {New York, NY},
  year      = {1983}
}

@book{bonnans2000perturbation,
  author    = {Bonnans, J. Fr{\'e}d{\'e}ric and Shapiro, Alexander},
  title     = {Perturbation Analysis of Optimization Problems},
  publisher = {Springer},
  address   = {New York, NY},
  year      = {2000}
}

@book{boucheron2013concentration,
  author    = {Boucheron, St{\'e}phane and Lugosi, G{\'a}bor and Massart, Pascal},
  title     = {Concentration Inequalities: A Nonasymptotic Theory of Independence},
  publisher = {Oxford University Press},
  address   = {Oxford},
  year      = {2013}
}

@incollection{mcdiarmid1989method,
  author    = {McDiarmid, Colin},
  title     = {On the Method of Bounded Differences},
  booktitle = {Surveys in Combinatorics},
  series    = {London Mathematical Society Lecture Note Series},
  volume    = {141},
  pages     = {148--188},
  publisher = {Cambridge University Press},
  year      = {1989}
}

@incollection{tsirelson1976norms,
  author    = {Tsirelson, Boris S. and Ibragimov, Ildar A. and Sudakov, Vladimir N.},
  title     = {Norms of {G}aussian Sample Functions},
  booktitle = {Proceedings of the Third Japan--USSR Symposium on Probability Theory},
  series    = {Lecture Notes in Mathematics},
  volume    = {550},
  pages     = {20--41},
  publisher = {Springer},
  year      = {1976}
}

@book{vandervaart1998asymptotic,
  author    = {van der Vaart, Aad W.},
  title     = {Asymptotic Statistics},
  publisher = {Cambridge University Press},
  address   = {Cambridge},
  year      = {1998}
}

@article{bousquet2002stability,
  author  = {Bousquet, Olivier and Elisseeff, Andr{\'e}},
  title   = {Stability and Generalization},
  journal = {Journal of Machine Learning Research},
  volume  = {2},
  pages   = {499--526},
  year    = {2002}
}
\bibliographystyle{iclr2027_conference}

\appendix
\section{Proofs}\label{app:proofs}

\paragraph{Proof of \cref{thm:concentration}.}
(i) Replacing one sample $c_i$ by $c_i'$ changes $\frac1n\sum_jv(c_j)$ by at most $D\norm{c_i-c_i'}/n\le D\delta/n$ (\cref{lem:value}). It moves $\bar c_n$ by $\norm{c_i-c_i'}/n$ and hence changes $v(\bar c_n)$ by at most $D\delta/n$. The bounded-differences inequality \citep{mcdiarmid1989method} with constants $2D\delta/n$ gives $2\exp(-2u^2/(n\cdot4D^2\delta^2/n^2))$.
(ii) Write $c_i=\mu+\Sigma^{1/2}\xi_i$ with $\xi_i\sim N(0,I)$ i.i.d., and view $\Regn$ as a function of $\xi=(\xi_1,\dots,\xi_n)\in\R^{nd}$. By \cref{lem:value} and Cauchy--Schwarz,
$|\frac1n\sum_i(v(c_i)-v(c_i'))|\le\frac Dn\opnorm{\Sigma}^{1/2}\sum_i\norm{\xi_i-\xi_i'}\le D\opnorm\Sigma^{1/2}n^{-1/2}\norm{\xi-\xi'}$.
Since $\norm{\bar\xi-\bar\xi'}\le n^{-1/2}\norm{\xi-\xi'}$, the same bound holds for $v(\bar c_n)$. So $\Regn$ is $L$-Lipschitz with $L=2D\opnorm\Sigma^{1/2}n^{-1/2}$, and Gaussian concentration \citep{tsirelson1976norms,boucheron2013concentration} gives $\P(|\Regn-\E\Regn|\ge u)\le2\exp(-u^2/(2L^2))$.
(iii) $\E\Regn-\Reg=\E v(\bar c_n)-v(\mu)$. This is $\le0$ by Jensen's inequality for the concave $v$, and $\ge-D\,\E\norm{\bar c_n-\mu}\ge-D\sqrt{\tr\Sigma/n}$ by \cref{lem:value} and Jensen's inequality. \qed

\paragraph{Proof of \cref{thm:clt}.}
A finite concave function that is differentiable at $\mu$ is Fr\'echet differentiable there, with gradient equal to its unique supergradient $\pis(\mu)$. Hence $v(\bar c_n)-v(\mu)=\pis(\mu)\T(\bar c_n-\mu)+o(\norm{\bar c_n-\mu})$. Since $\sqrt n\norm{\bar c_n-\mu}=O_P(1)$,
\[
\sqrt n(\Regn-\Reg)=\frac1{\sqrt n}\sum_{i=1}^n\Bigl[\pis(\mu)\T(c_i-\mu)-\bigl(v(c_i)-\E v(c)\bigr)\Bigr]+o_P(1).
\]
The summands are i.i.d.\ with mean zero and variance $\Var(c\T\pis(\mu)-v(c))=\Var(\ell(\mu,c))$, which is finite because $v$ is Lipschitz and $\E\norm c^2<\infty$. The claim follows from the CLT and Slutsky's lemma \citep{vandervaart1998asymptotic}. At a kink, the directional derivative of $v$ at $\mu$ in direction $d$ is $\min_{w\in\partial v(\mu)}w\T d$ \citep[Thm.~23.4]{rockafellar1970convex}, which yields the non-Gaussian limit stated after the theorem. \qed

\section{Quadratic programs}\label{app:qp}
\paragraph{(i) Unconstrained.} The first-order condition $c+\lambda Qz=0$ gives $\pis(c)=-\lambda^{-1}Q^{-1}c$, so $A=-\lambda^{-1}Q^{-1}$ in \cref{thm:affine}, and $\Reg=\tr(Q^{-1}\Sigma)/(2\lambda)$. Directly, $v(c)=-\frac1{2\lambda}c\T Q^{-1}c$, and $v(\mu)-\E v(c)=\frac1{2\lambda}\tr(Q^{-1}\Sigma)$. The growth condition \cref{ass:growth}(b) holds with $a=0$.

\paragraph{(ii) Equality constraints.} The KKT system $c+\lambda Qz+A\T\nu=0$, $Az=b$ has the unique solution $\nu=-(AQ^{-1}A\T)^{-1}(\lambda b+AQ^{-1}c)$ and $\pis(c)=Mc+m_0$ with $M$ as stated and $m_0=Q^{-1}A\T(AQ^{-1}A\T)^{-1}b$. $M$ is symmetric and negative semidefinite, so $-\frac12\tr(M\Sigma)\ge0$.

\paragraph{(iii) Inequality constraints.} Completing the square shows that $\pis(c)$ is the $Q$-weighted projection of $-\lambda^{-1}Q^{-1}c$ onto $Z$. Projections onto closed convex sets are nonexpansive in the corresponding norm, so $\pis$ is Lipschitz. Under LICQ and strict complementarity the active set is locally constant, and $\pis$ is affine on each active-set region \citep{fiacco1983introduction,bonnans2000perturbation}.

\section{Experimental details}\label{app:details}
\paragraph{Synthetic problems (\cref{sec:exp-synthetic}).} The problems are: an LP over the simplex; an LP over $[-1,1]^d$ with cost $c-\tau$ and $\tau$ the population mean, so that the linear shift enters as $h(z)=-\tau\T z$; an LP over the convex hull of 40 Gaussian points; the $5\times5$ monotone grid shortest path; an unconstrained QP; a budget-constrained QP; and a simplex-constrained QP with diagonal $Q$, solved exactly by bisection on the budget multiplier. The costs are correlated Gaussian, uniform, Student-$t_5$, log-normal, gamma, Poisson, a two-component Gaussian mixture, and coordinate-wise two-point. All quantities are computed under the empirical distribution of $n=20{,}000$ draws, with $\mu$ set to the sample mean; standard errors against the population values are computed by batch means and reported in the released CSV.

\paragraph{Shortest path (\cref{sec:exp-eg}).} All 70 monotone paths are enumerated, so $\pis$ is an exact argmin over path costs. A single $B^\star$ is shared across degrees. The least-squares predictor includes an intercept and is fitted on an independent training set of 1{,}000 draws for each degree. The Bayes predictor uses the true conditional mean $(p^{-1/2}B^\star x+3)^{\deg}+1$, since $\E[\varepsilon_j]=1$.

\section{Radius $D$ for common feasible sets}\label{app:examples}
\Cref{thm:bound} needs only $D=\sup_{z\in Z}\norm z$, which is often available by inspection.

\begin{center}
\small
\begin{tabular}{ll}
\toprule
Feasible set & $D$ \\
\midrule
Probability simplex (portfolio, budget and no short sales) & $1$ \\
$k\times k$ monotone grid paths & $\sqrt{2(k-1)}$ \\
Assignment / Birkhoff polytope, $m\times m$ & $\sqrt m$ \\
Spanning trees on $m$ nodes & $\sqrt{m-1}$ \\
Tours on $m$ cities & $\sqrt m$ \\
$\ell_1$ ball of radius $t$; $\ell_2$ ball of radius $t$ & $t$ \\
Box $[0,u]^d$ & $\norm u$ \\
\bottomrule
\end{tabular}
\end{center}

\section{Additional figures}\label{app:figures}

\begin{figure}[h]
\centering
\includegraphics[width=\linewidth]{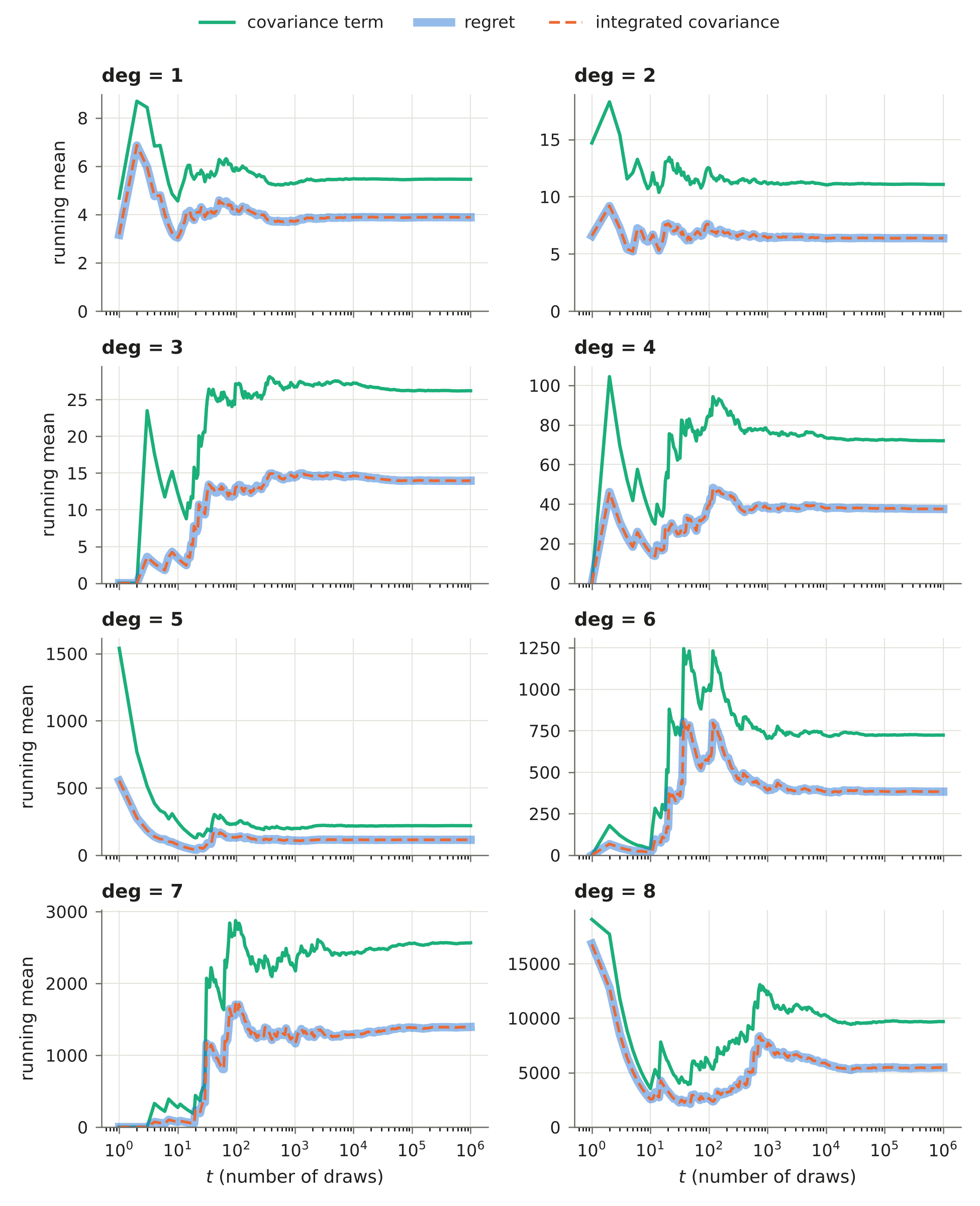}
\caption{Running means over $t=1,\dots,10^6$ draws for the Bayes predictor. Regret and integrated covariance coincide at every $t$, including $t=1$, because \cref{thm:pathwise} holds scenario by scenario. The covariance term lies above both.}
\label{fig:convergence}
\end{figure}

\end{document}